\documentclass{cernrep}
\newcommand{\babar}{\mbox{\slshape B\kern-0.1em{\small A}\kern-0.1em B\kern-0.1em{\small A\kern-0.2em R}}}
\newcommand{\E}{\mathrm{E}}
\newcommand{\C}{\mathrm{C}}
\newcommand{\dd}[2]{\frac{\partial{#1}}{\partial{#2}}}

\title{Unfolding algorithms and tests using RooUnfold}
\author{Tim Adye}
\institute{Particle Physics Department, Rutherford Appleton Laboratory, Didcot, United Kingdom.}
\begin{document}
\maketitle

\begin{abstract}
The RooUnfold package provides a common framework to evaluate and use
different unfolding algorithms, side-by-side. It currently provides
implementations or interfaces for the Iterative Bayes, Singular Value Decomposition, and
TUnfold methods, as well as bin-by-bin and matrix inversion reference
methods. Common tools provide covariance matrix evaluation and
multi-dimensional unfolding. A test suite allows comparisons of the
performance of the algorithms under different truth and measurement models.
Here I outline the package, the unfolding methods, and some experience of their use.
\end{abstract}

\section{RooUnfold package aims and features}

The RooUnfold package~\cite{RooUnfold-web} was designed to provide a framework for different unfolding algorithms.
This approach simplifies the comparison between algorithms and has allowed
common utilities to be written.
Currently RooUnfold implements or interfaces to the Iterative Bayes~\cite{D'Agostini:1994zf,Bierwagen:PHYSTAT2011},
Singular Value Decomposition (SVD)~\cite{Hocker:1995kb,Kartvelishvili:PHYSTAT2011,Tackmann:PHYSTAT2011},
TUnfold~\cite{Schmitt-web}, bin-by-bin correction factors, and unregularized matrix inversion methods.

The package is designed around a simple object-oriented approach, implemented in
C++, and using existing ROOT~\cite{Brun:1997pa} classes. RooUnfold defines classes for the different
unfolding algorithms, which inherit from a common base class, and a class for
the response matrix. The response matrix object is independent of the unfolding,
so can be filled in a separate `training' program.

RooUnfold can be linked into a stand-alone program, run from a ROOT/CINT script, or
executed interactively from the ROOT prompt.
The response matrix can be initialized using existing histograms or matrices, or
filled with built-in methods (these can take care of the normalization when inefficiencies are to be considered).
The results can be returned as a histogram with errors, or a vector with full covariance matrix.
The framework also takes care of handling multi-dimensional distributions
(with ROOT support for 1--, 2--, and 3--dimensional (1D,2D,3D) histograms),
different binning for measured and truth distributions,
variable binning, and the option to include or exclude under- and over-flows.
It also supports different methods for calculating the errors that can
be selected with a simple switch: bin-by-bin errors with no correlations,
the full covariance matrix from the propagation of measurement errors in the unfolding, or
the covariance matrix calculated using Monte Carlo (MC) toys.

All these details are handled by the framework, so do not have to be
implemented for each algorithm. However different bin layouts may not produce good results for
algorithms that rely on the global shape of the distribution (SVD).

A toy MC test framework is provided, allowing
selection of different MC probability density functions (PDF) and parameters,
comparing different binning, and performing the unfolding with the different
algorithms and varying the unfolding regularization parameters.
Tests can be performed with 1D, 2D, and 3D distributions.
The results of a few such tests are presented in section~\ref{sec:adye:examples}.

\section{C++ classes}

\begin{figure}
\centerline{\includegraphics[width=0.8\textwidth]{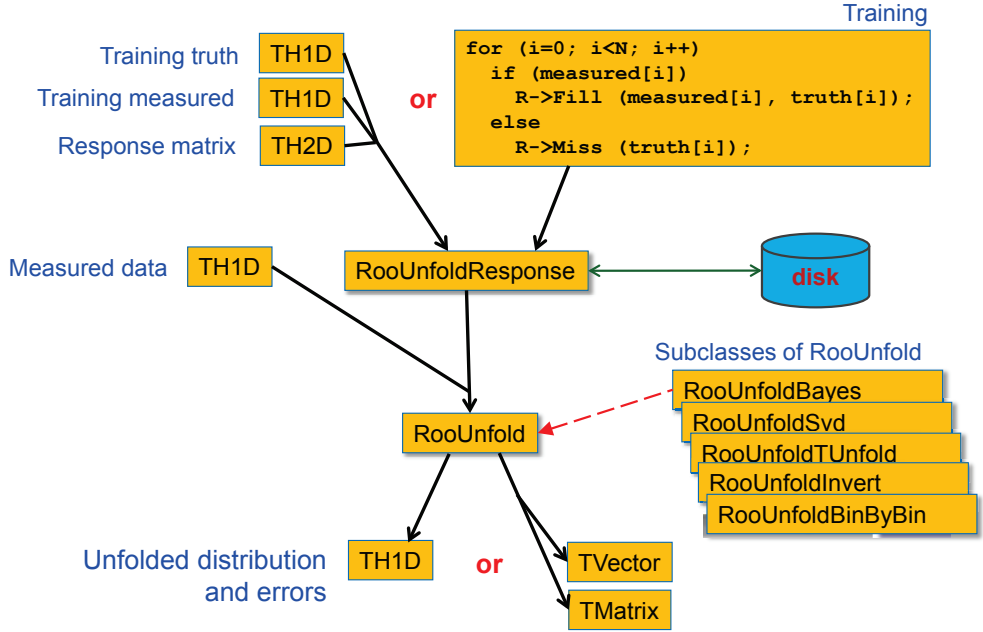}}
\caption
[The RooUnfold classes.]%
{The RooUnfold classes. The training truth, training measured, measured data, and unfolded distributions
can also be given as TH2D or TH3D histograms.}%
\label{Fig:adye:classes}%
\end{figure}

Figure~\ref{Fig:adye:classes} summarizes how the ROOT and RooUnfold classes are used
together. The RooUnfoldResponse object can be constructed using a 2D response histogram (TH2D)
and 1D truth and measured projections (these are required to determine the effect of inefficiencies).
Alternatively, RooUnfoldResponse can be filled directly with the
\texttt{Fill($x_{\rm measured}$, $x_{\rm true}$)}
and
\texttt{Miss($x_{\rm true}$)}
methods, where the \texttt{Miss} method is used to count an event that was not measured
and should be counted towards the inefficiency.%

The RooUnfoldResponse object can be saved to disk using the usual ROOT input/output
streamers. This allows the easy separation in separate programs
of MC training from the unfolding step.

A RooUnfold object is constructed using a RooUnfoldResponse object and the measured
data. It can be constructed as a RooUnfoldBayes, RooUnfoldSvd, RooUnfoldTUnfold, (etc)
object, depending on the algorithm required.

The results of the unfolding can be obtained as ROOT histograms (TH1D, TH2D, or TH3D)
or as a ROOT vector (TVectorD) and covariance matrix (TMatrixD). The histogram will
include just the diagonal elements of the error matrix. This should be used with care,
given the significant correlations that can occur if there is much bin-to-bin migration.

\section{Unfolding algorithms}

\subsection{Iterative Bayes' theorem}

The RooUnfoldBayes algorithm uses the method described by D'Agostini in~\cite{D'Agostini:1994zf}.
Repeated application of Bayes' theorem is used to invert the response matrix.
Regularization is achieved by stopping iterations before reaching the `true'
(but wildly fluctuating) inverse.
The regularization parameter is just the number of iterations.
In principle, this has to be tuned according to the sample statistics and binning.
In practice, the results are fairly insensitive to the precise setting used
and four iterations are usually sufficient.

RooUnfoldBayes takes the training truth as its initial prior, rather than a flat distribution,
as described by D'Agostini.
This should not bias result once we have iterated, but could reach an optimum after fewer iterations.

This implementation takes account of errors on the data sample but not,
by default, uncertainties in the response matrix due to finite MC statistics.
That calculation can be very slow, and usually the training sample is much larger
than the data sample.

RooUnfoldBayes does not normally do smoothing, since this has not been found to be necessary
and can, in principle, bias the distribution. Smoothing can be enabled with an option.

\subsection{Singular Value Decomposition}

RooUnfoldSvd provides an interface to the
TSVDUnfold class implemented in ROOT by Tackmann~\cite{Tackmann:PHYSTAT2011}, which
uses the method of H\"ocker and Kartvelishvili~\cite{Hocker:1995kb}.
The response matrix is inverted using singular value decomposition,
which allows for a linear implementation of the unfolding algorithm.
The normalization to the number of events is retained in order to minimize
uncertainties due to the size of the training sample.
Regularization is performed using a smooth cut-off on small singular value contributions
($s_i^2 \rightarrow s_i^2 / (s_i^2 + s_k^2)$, where the $k$th singular value defines the cut-off),
which correspond to high-frequency fluctuations.

The regularization needs to be tuned according to the distribution, binning, and sample statistics
in order minimize the bias due to the choice of the training sample (which dominates at small $k$)
while retaining small statistical fluctuations in the unfolding result (which grow at large $k$).

The unfolded error matrix includes the contribution of uncertainties on the
response matrix due to finite MC training statistics.

\subsection{TUnfold}

RooUnfoldTUnfold provides an interface to the TUnfold method implemented in ROOT by Schmitt~\cite{Schmitt-web}.
TUnfold performs a matrix inversion with 0-, 1-, or 2-order polynomial regularization of neighbouring bins.
RooUnfold automatically takes care of packing 2D and 3D distributions
and creating the appropriate regularization matrix required by TUnfold.

TUnfold can automatically determine an optimal regularization parameter ($\tau$) by scanning the
`L-curve' of $\log_{10} \chi^2$ vs $\log_{10} \tau$.

\subsection{Unregularized algorithms}

Two simple algorithms,
RooUnfoldBinByBin, which applies MC correction factors with no inter-bin migration,
and RooUnfoldInvert, which performs unregularized matrix inversion with singular value removal (TDecompSVD)
are included for reference.
These methods are not generally recommended: the former risks biases from the MC model,
while the latter can give large bin-bin correlations and magnify statistical fluctuations.

\section{Examples\label{sec:adye:examples}}

\begin{figure}
\makebox[\textwidth]{\includegraphics[angle=-90,width=.640\textwidth,clip]{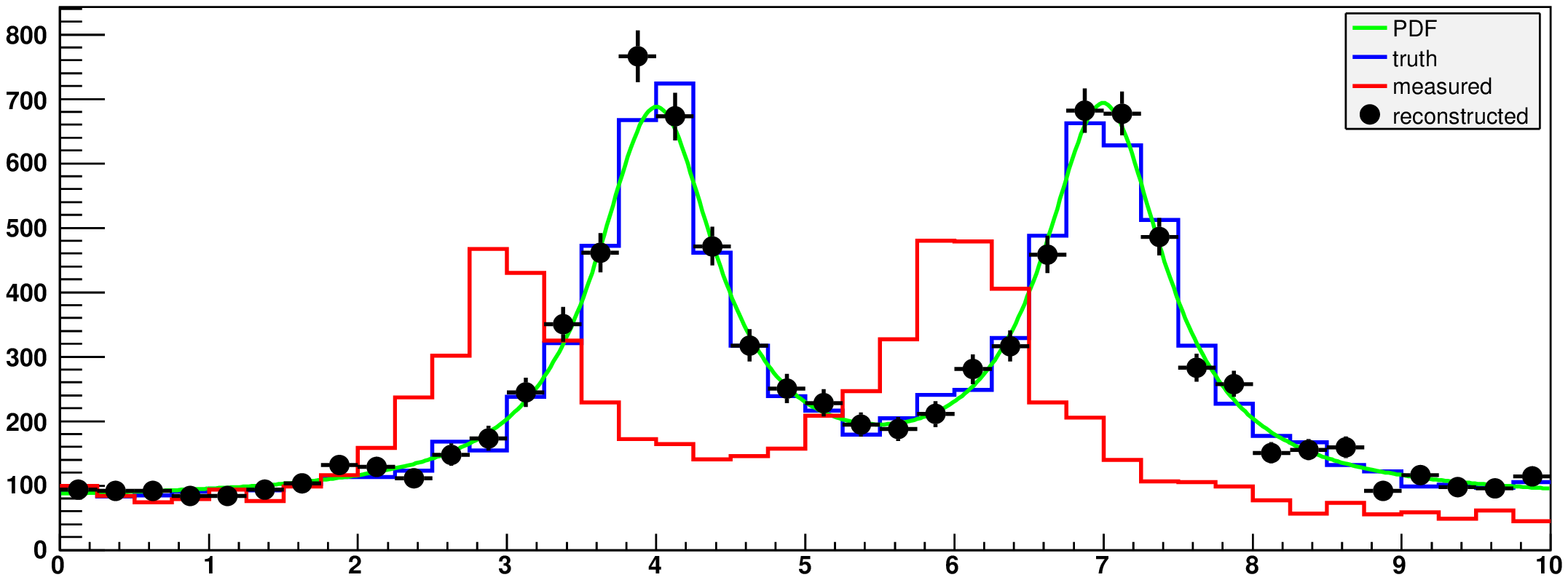}\hfill
                     \includegraphics[angle=-90,width=.338\textwidth,clip]{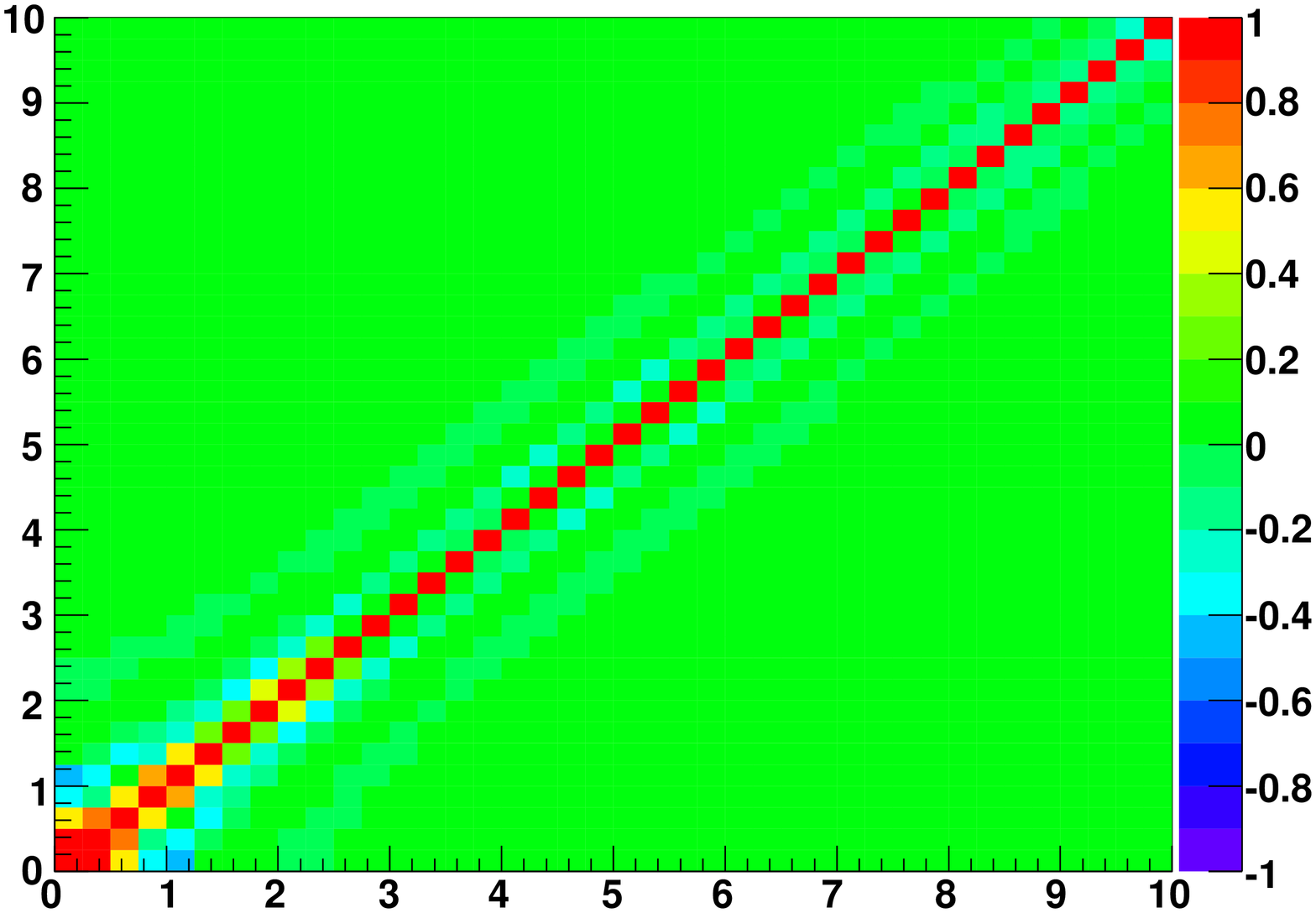}}%
\caption
[Unfolding with the Bayes algorithm.]%
{Unfolding with the Bayes algorithm.
On the left, a double Breit-Wigner PDF on a flat background (green curve) is used to generate
a test `truth' sample (upper histogram in blue).
This is then smeared, shifted, and a variable inefficiency applied to produce
the `measured' distribution (lower histogram in red).
Applying the Bayes algorithm with 4~iterations on this latter gave the unfolded result
(black points), shown with errors from the diagonal elements of the error matrix.
The bin-to-bin correlations from the error matrix are shown on the right.}%
\label{Fig:adye:bayes-example}%
\end{figure}

\begin{figure}
\makebox[\textwidth]{\includegraphics[angle=-90,width=.640\textwidth,clip]{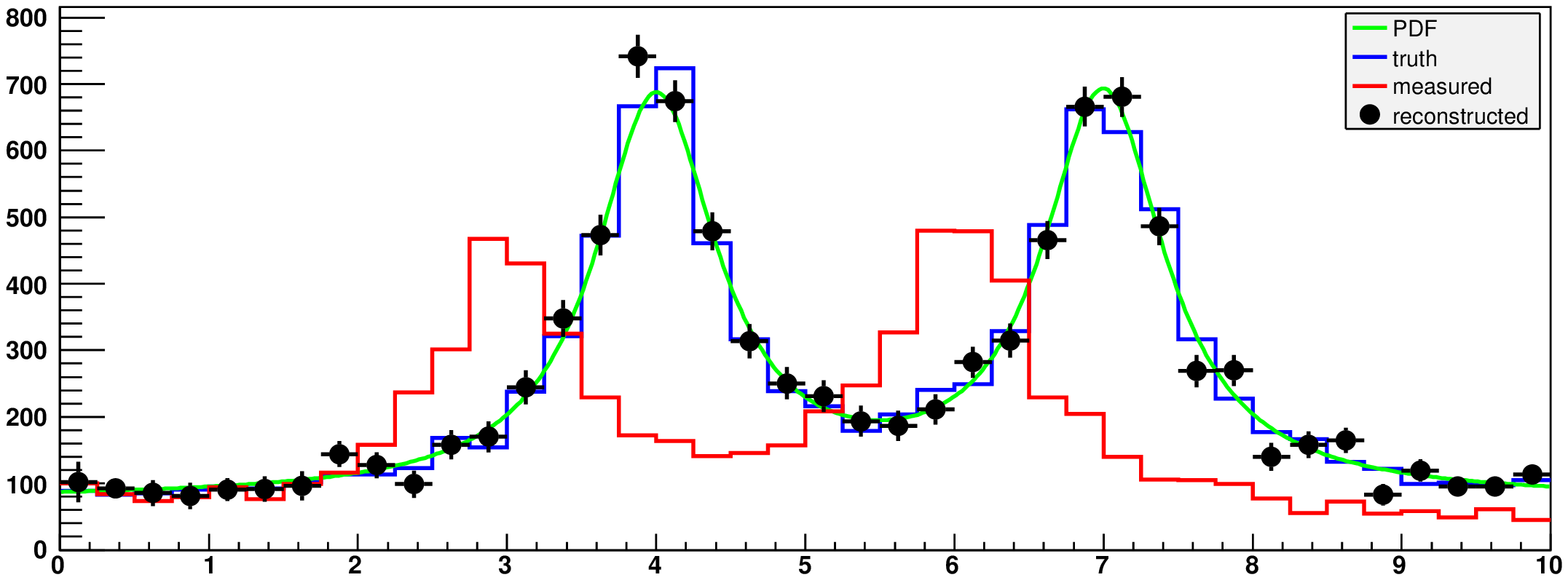}\hfill
                     \includegraphics[angle=-90,width=.338\textwidth,clip]{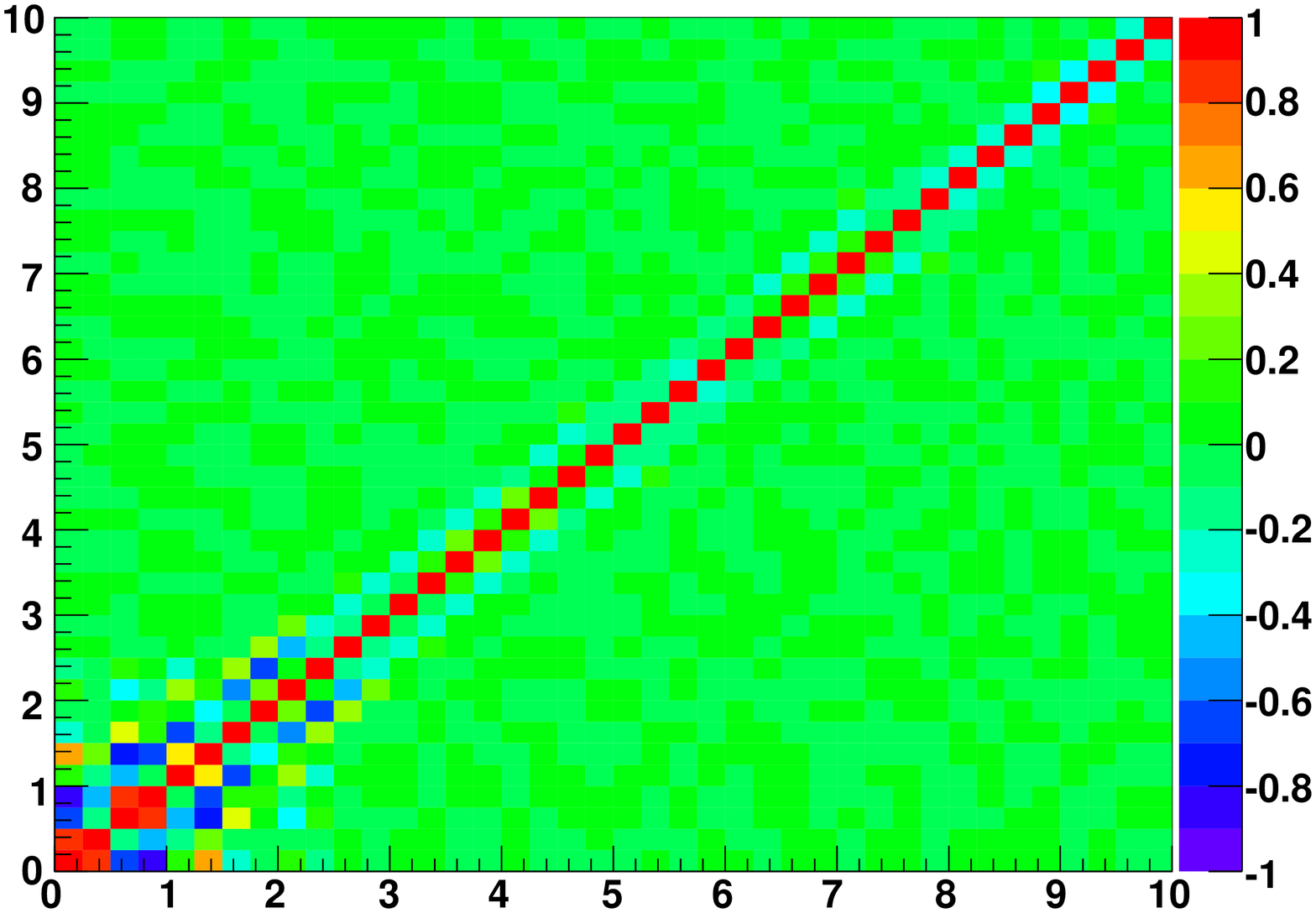}}%
\caption
[Unfolding with the SVD algorithm]%
{Unfolding with the SVD algorithm ($k=30$) on the same training and test
samples as described in Fig.~\ref{Fig:adye:bayes-example}.}%
\label{Fig:adye:svd-example}%
\end{figure}

\begin{figure}
\makebox[\textwidth]{\includegraphics[angle=-90,width=.640\textwidth,clip]{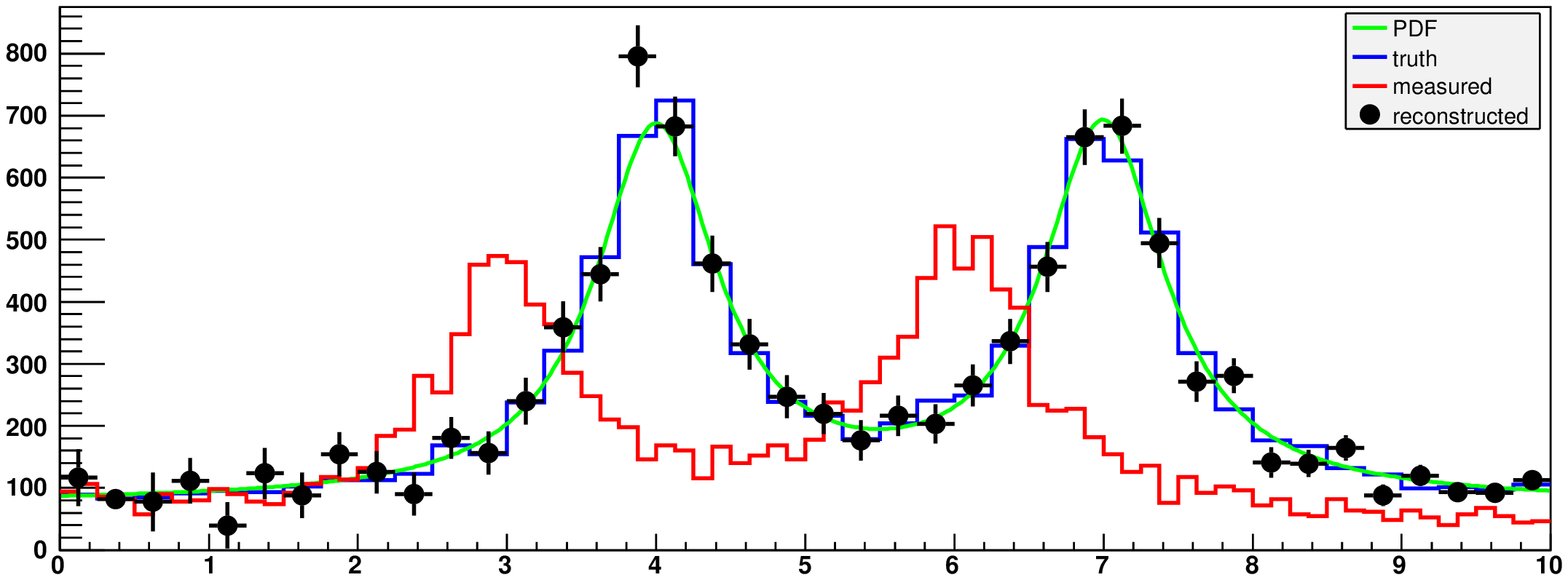}\hfill
                     \includegraphics[angle=-90,width=.338\textwidth,clip]{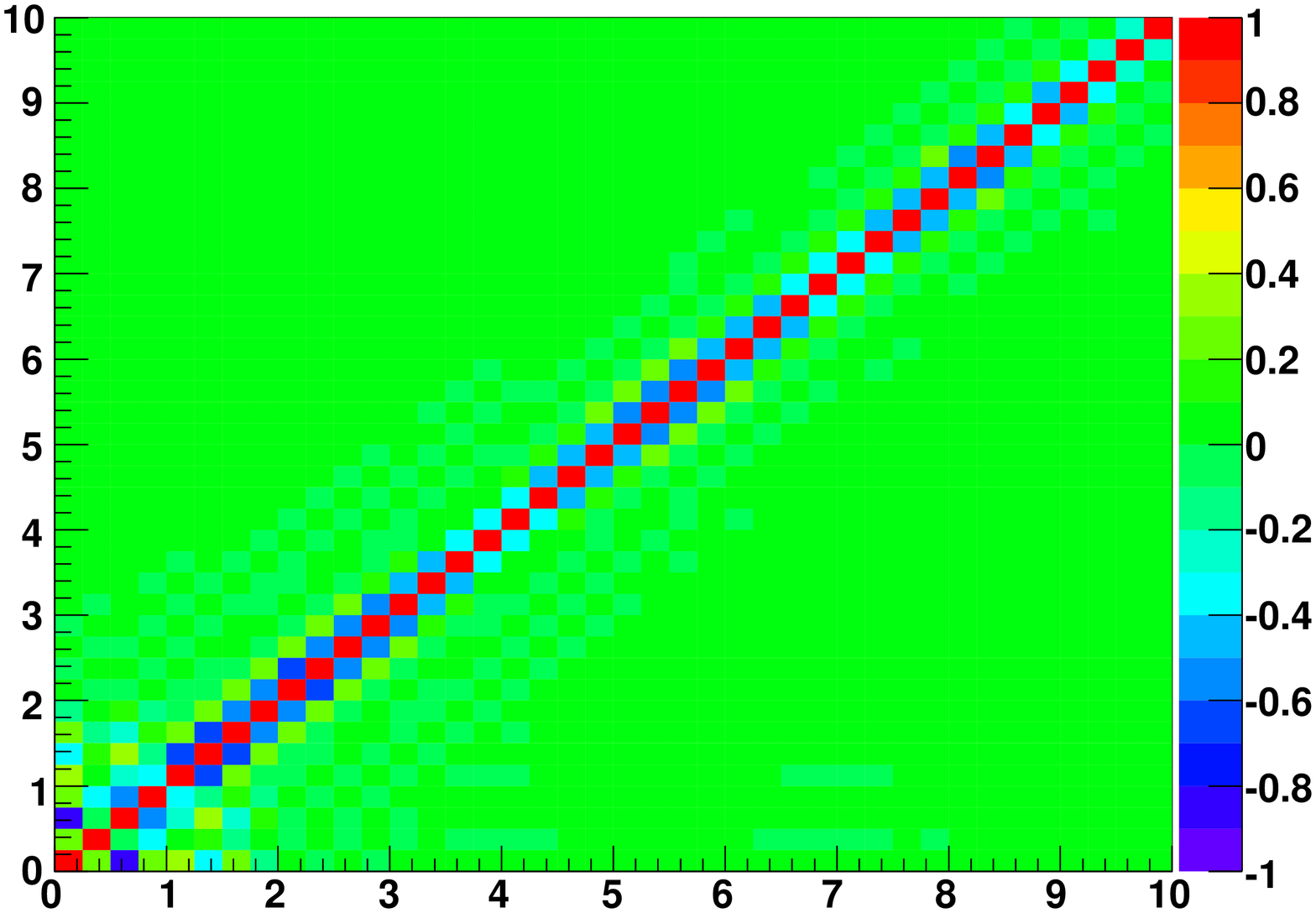}}%
\caption
[Unfolding with the TUnfold algorithm]%
{Unfolding with the TUnfold algorithm ($\tau=0.004$) on the same training and test
samples as described in Fig.~\ref{Fig:adye:bayes-example}.
Here we use two measurement bins for each truth bin.}%
\label{Fig:adye:tunfold-example}
\end{figure}

Examples of toy MC tests generated by RooUnfoldTest
are shown in Figs.~\ref{Fig:adye:bayes-example}--\ref{Fig:adye:tunfold-example}.
These provide a challenging test of the procedure.
Completely different training and test MC models are used:
a single wide Gaussian PDF for training and a
double Breit-Wigner for testing. In both cases
these are smeared, shifted, and a variable inefficiency
applied to produce the `measured' distributions.%

\section{Unfolding errors}

Regularization introduces inevitable correlations between bins in the unfolded distribution.
To calculate a correct $\chi^2$, one has to invert the covariance matrix:
\begin{equation}
\chi^2 = (\mathbf{x}_{\mathrm{measured}} - \mathbf{x}_{\mathrm{true}})^{\mathrm{T}} \mathbf{V}^{-1}
         (\mathbf{x}_{\mathrm{measured}} - \mathbf{x}_{\mathrm{true}})
\end{equation}

However, in many cases, the covariance matrix is poorly conditioned,
which makes calculating the inverse problematic.
Inverting a poorly conditioned matrix involves subtracting large, but
very similar numbers, leading to significant effects due to the
machine precision.

\subsection{Unfolding errors with the Bayes method}

\begin{figure}
\makebox[\textwidth]{\includegraphics[width=.47\textwidth,clip]{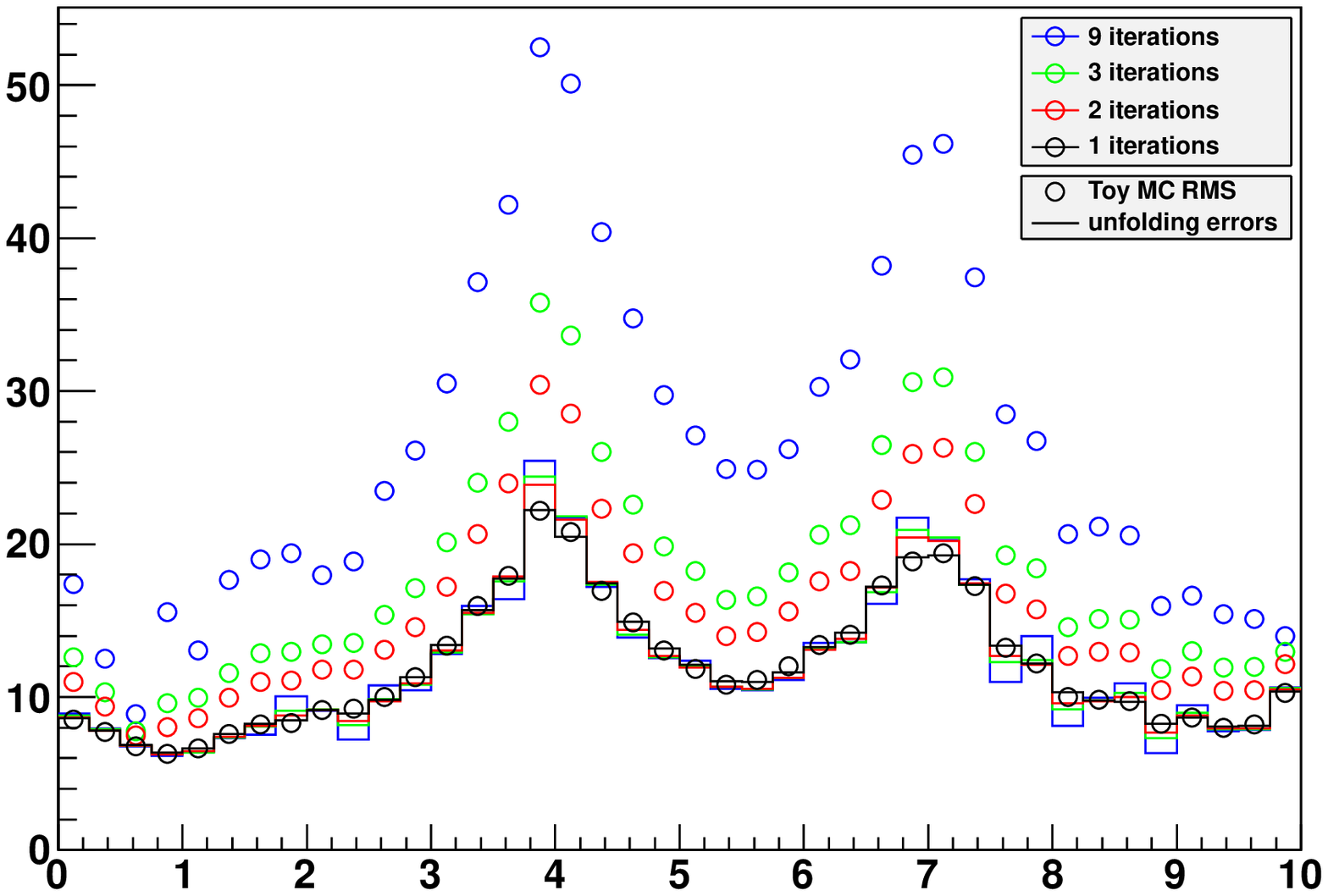}\hfill
                     \includegraphics[width=.47\textwidth,clip]{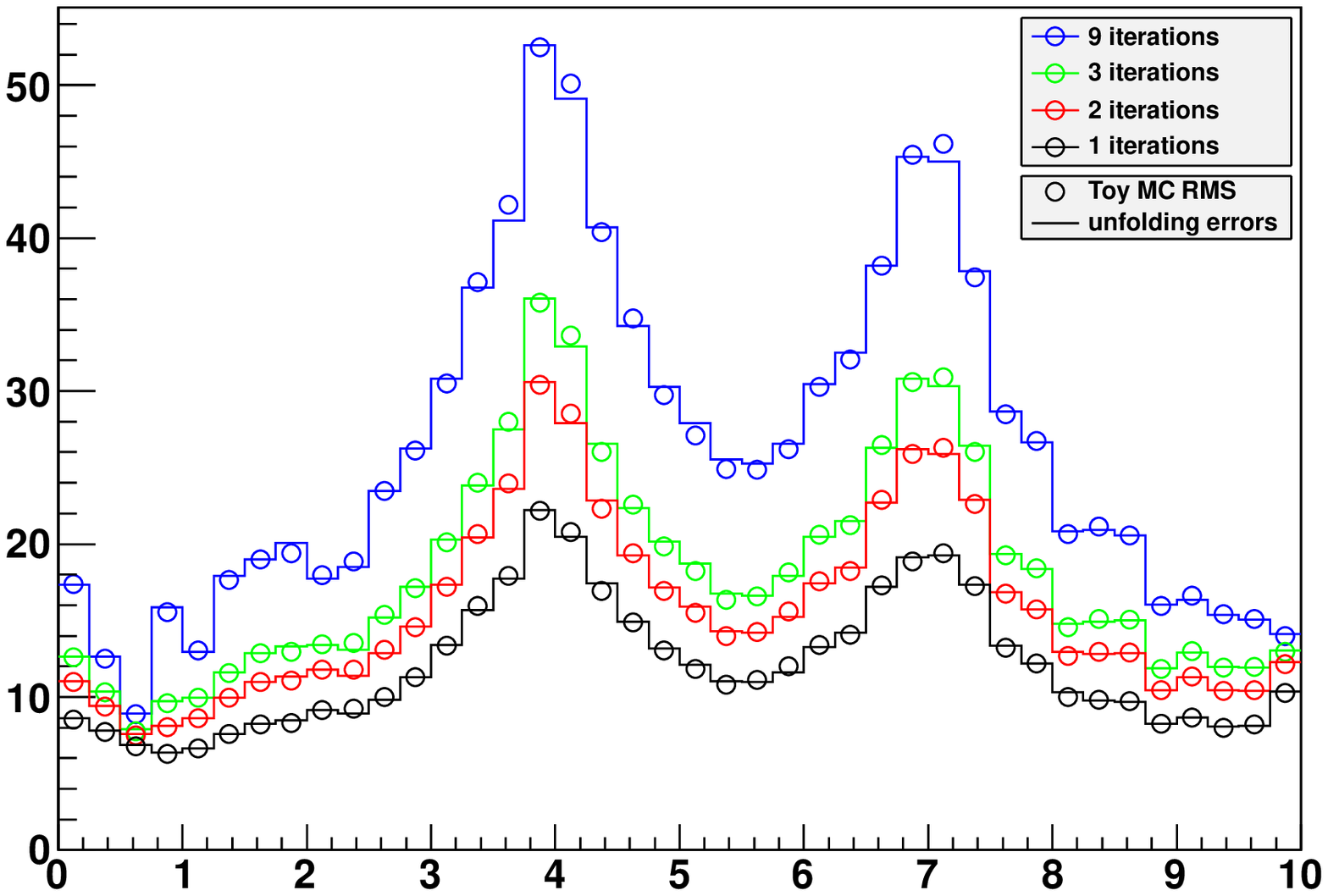}}%
\caption
[Bayesian unfolding errors compared to toy MC]%
{Bayesian unfolding errors (lines) compared to toy MC RMS (points) for 1, 2, 3, and 9 iterations
on the Fig.~\ref{Fig:adye:bayes-example} test.
The left-hand plot shows the errors using D'Agostini's original method,
ignoring any dependence on previous iterations (only the $M_{ij}$ term in Eq.~(\ref{eq:dnCidnEj})).
The right-hand plot shows the full error propagation.}%
\label{fig:bayes_errors}%
\end{figure}

As shown on the left-hand side of Fig.~\ref{fig:bayes_errors},
the uncertainties calculated by propagation of errors in the Bayes method
were found to be significantly underestimated compared to those given by the toy MC.
This was found to be due to an omission in the original method
outlined by D'Agostini (\cite{D'Agostini:1994zf}~section~4).%

The Bayes method gives the unfolded distribution (`estimated causes'), $\hat{n}(\C_i)$,
as the result of applying the unfolding matrix, $M_{ij}$, to the measurements (`effects'), $n(\E_j)$:
\begin{equation}
\hat{n}(\C_i) = \sum_{j=1}^{n_{\E}} M_{ij} n(\E_j)
\quad\mathrm{where}\quad
M_{ij} = \frac{P(\E_j|\C_i) n_0(\C_i)}{\epsilon_i \sum_{l=1}^{n_{\C}} P(\E_j|\C_l) n_0(C_l)}
\label{eq:nCi}
\end{equation}
\noindent $P(\E_j|\C_i)$ is the $n_{\E} \times n_{\C}$ response matrix,
$\epsilon_i \equiv \sum_{j=1}^{n_{\E}} P(\E_j|\C_i)$ are efficiencies, and
$n_0(C_l)$ is the prior distribution --- initially arbitrary (eg. flat or MC model), but updated on
subsequent iterations.

The covariance matrix, which here we call $V(\hat{n}(\C_k),\hat{n}(\C_l))$,
is calculated by error propagation from $n(\E_j)$,
but $M_{ij}$ is assumed to be itself independent of $n(\E_j)$. That is only true for the first iteration.
For subsequent iterations, $n_0(\C_i)$ is replaced by $\hat{n}(\C_i)$ from the
previous iteration, and $\hat{n}(\C_i)$ depends on $n(\E_j)$ (Eq.~(\ref{eq:nCi})).

To take this into account, we compute the error propagation matrix
\begin{equation}
\dd{\hat{n}(\C_i)}{n(\E_j)} = M_{ij} + \sum_{k=1}^{n_{\E}} M_{ik} n(\E_k)
\left( \frac{1}{n_0(\C_i)} \dd{n_0(\C_i)}{n(\E_j)} - \sum_{l=1}^{n_{\C}} \frac{\epsilon_l}{n_0(\C_l)} \dd{n_0(\C_l)}{n(\E_j)} M_{lk} \right)
\label{eq:dnCidnEj}
\end{equation}
This depends upon the matrix $\dd{n_0(\C_i)}{n(\E_j)}$, which is $\dd{\hat{n}(\C_i)}{n(\E_j)}$ from the previous iteration.
In the first iteration, the second term vanishes ($\dd{n_0(\C_i)}{n(\E_j)}=0$) and we get $\dd{\hat{n}(\C_i)}{n(\E_j)} = M_{ij}$.

The error propagation matrix can be used to obtain the covariance matrix on the unfolded distribution
\begin{equation}
V(\hat{n}(\C_k),\hat{n}(\C_l)) = \sum_{i,j=1}^{n_{\E}} \dd{\hat{n}(\C_k)}{n(\E_i)} V(n(\E_i),n(\E_j)) \dd{\hat{n}(\C_l)}{n(\E_j)}
\label{eq:Vij}
\end{equation}
\noindent from the covariance matrix of the measurements, $V(n(\E_i),n(\E_j))$.

Without the new second term in Eq.~(\ref{eq:dnCidnEj}),
the error is underestimated if more than one iteration
is used, but agrees well with toy MC tests if the full error propagation is used,
as shown in Fig.~\ref{fig:bayes_errors}.%

\section{Status and plans}

RooUnfold was first developed in the \babar\ software environment and
released stand-alone in 2007.
Since then, it has been used by physicists from many
different particle physics, particle-astrophysics, and nuclear physics groups.
Questions, suggestions, and bug reports from users
have prompted new versions with fixes and improvements.

Last year I started working with a small group hosted by the
Helmholtz Alliance, the Unfolding Framework Project\cite{unfolding-project}.
The project is developing unfolding experience, software, algorithms, and performance tests.
It has adopted RooUnfold as a framework for development.

Development and improvement of RooUnfold is continuing.
In particular, determination of the systematic errors due to uncertainties on
the response matrix, and due to correlated measurement bins will be added.
The RooUnfold package will be incorporated into the ROOT distribution,
alongside the existing TUnfold and TSVDUnfold classes.



\begin{thebibliography}{99}

\bibitem{RooUnfold-web}
  The RooUnfold package and documentation are available from\\
  \verb=http://hepunx.rl.ac.uk/~adye/software/unfold/RooUnfold.html=

\bibitem{D'Agostini:1994zf}
  G.~D'Agostini,
  ``A Multidimensional unfolding method based on Bayes' theorem'',
  Nucl.\ Instrum.\ Meth.\  A {\bf 362} (1995) 487.

\bibitem{Bierwagen:PHYSTAT2011}
  K.~Bierwagen,
  ``Bayesian Unfolding'',
  presented at PHYSTAT 2011 (CERN, Geneva, January 2011), to be published in a CERN Yellow Report.

\bibitem{Hocker:1995kb}
  A.~Hocker and V.~Kartvelishvili,
  ``SVD Approach to Data Unfolding'',
  Nucl.\ Instrum.\ Meth.\  A {\bf 372} (1996) 469.

\bibitem{Kartvelishvili:PHYSTAT2011}
  V.~Kartvelishvili,
  ``Unfolding with SVD'',
  presented at PHYSTAT 2011 (CERN, Geneva, January 2011), to be published in a CERN Yellow Report.

\bibitem{Tackmann:PHYSTAT2011}
  K.~Tackmann,
  ``SVD-based unfolding: implementation and experience'',
  presented at PHYSTAT 2011 (CERN, Geneva, January 2011), to be published in a CERN Yellow Report.

\bibitem{Schmitt-web}
  The TUnfold package is available in ROOT~\cite{Brun:1997pa} and documented in\\
  \verb=http://www.desy.de/~sschmitt/tunfold.html=

\bibitem{Brun:1997pa}
  R.~Brun and F.~Rademakers,
  ``ROOT: An object oriented data analysis framework'',
  Nucl.\ Instrum.\ Meth.\  A {\bf 389} (1997) 81.
  See also \verb=http://root.cern.ch/=.

\bibitem{unfolding-project}
  For details of the Unfolding Framework Project, see\\
  \verb=https://www.wiki.terascale.de/index.php/Unfolding_Framework_Project=

\end{thebibliography}
\end{document}